\documentclass{article}

\usepackage{graphicx}
\usepackage{fullpage,amsmath,amssymb,latexsym}
\usepackage{multirow}
\usepackage{natbib}
\begin{document}
\title{Uranus and Neptune: Shape and Rotation}
\author{Ravit Helled$^1$, John D. Anderson$^2$, and  Gerald Schubert$^1$\\
\small{$^1$Department of Earth and Space Sciences and Institute of Geophysics and Planetary Physics,}\\
\small{University of California, Los Angeles, CA 90095Ð1567, USA}\\
\small{$^2$Jet Propulsion Laboratory \footnote{Retiree},} \\
\small{California Institute of Technology, Pasadena, CA 91109}\\
\small{corresponding email: rhelled@ucla.edu (R. Helled)}\\
%jdandy@earthlink.net (J. D. Anderson); 
%schubert@ucla.edu (G. Schubert)}\\
}
\date{}
\maketitle 

\begin{abstract}
Both Uranus and Neptune are thought to have strong zonal winds with velocities of several hundred meters per second. These wind velocities, however, assume solid-body rotation periods based on Voyager 2 measurements of periodic variations in the planets' radio signals and of fits to the planets' magnetic fields; 17.24h and 16.11h for Uranus and Neptune, respectively. The realization that the radio period of Saturn does not represent the planet's deep interior rotation and the complexity of the magnetic fields of Uranus and Neptune raise the possibility that the Voyager 2 radio and magnetic periods might not represent the deep interior rotation periods of the ice giants. Moreover, if there is deep differential rotation within Uranus and Neptune no single solid-body rotation period could characterize the bulk rotation of the planets. We use wind and shape data to investigate the rotation of Uranus and Neptune. 
The shapes (flattening) of the ice giants are not measured, but only inferred from atmospheric wind speeds and radio occultation measurements at a single latitude. The inferred oblateness values of Uranus and Neptune do not correspond to bodies rotating with the Voyager rotation periods. 
Minimization of wind velocities or dynamic heights of the 1 bar isosurfaces, constrained by the single occultation radii and gravitational coefficients of the planets, leads to solid-body rotation periods of $\sim$ 16.58h for Uranus and $\sim$ 17.46h for Neptune. Uranus might be rotating faster and Neptune slower than Voyager rotation speeds. We derive shapes for the planets based on these rotation rates. Wind velocities with respect to these rotation periods are essentially identical on Uranus and Neptune and wind speeds are slower than previously thought. 
Alternatively, if we interpret  wind measurements in terms of differential rotation on cylinders there are essentially no residual atmospheric winds. 
 
\end{abstract}

\newpage

\section{Introduction}
In this paper we review what is regarded as known about the shapes and rotation rates of the icy giants Uranus and Neptune. Though the literature provides values of the planets' rotation periods and flattening we point out that the reported shapes do not correspond to the Voyager rotation periods. The overall shape of a rotating planet in hydrostatic equilibrium is predictable from its rotation period and internal structure, as revealed by its gravitational coefficients, and it is in this sense that generally accepted values of period, oblateness, and gravitational field do not agree for the icy giants. \par

We explore the relation between shape and rotation and point out that our knowledge of the shapes of the icy giants is not as firm as might be thought. Voyager radio occultations were limited and did not measure the actual shapes. Stellar occultations of the icy giants provide information about their upper atmospheres but their downward extrapolations to the lower atmosphere depend on unknown zonal wind velocity distributions.\par

With regard to our knowledge of icy planet rotation periods, it must be acknowledged that there is uncertainty because we use the radio signals and magnetic fields as proxies of the planetary bulk rotation. 
Periods inferred from the radio signals and magnetic fields might be equal if the radiation emanates from charged particles simply attached to the magnetic field lines, but the periods could be different if the radiation originates from a small concentration of ions in a centrifugally loaded magnetosphere. There is also the question of what parts of the interior of a planet are tied to the magnetic field lines. The magnetic fields of Uranus and Neptune are strongly multi-polar and may originate at shallow depths in the planets. If the region interior to the dynamo generation layer is not as electrically conducting as the dynamo region (perhaps due to the stratification of ices and silicates) then the deep interior could rotate differentially with respect to the dynamo layer.
In this case the radio emissions or the magnetic fields might not reveal the rotation of the bulk of the planet. 
Are the radio periods constant with time? We have seen that Saturn's radio emissions have multiple periodicities and vary in time (Gurnett et al., 2009). The Voyager observations are the only ones we have of the icy giants' radio signals and we are unable to assess the stability of the radio periodicities. Not to be left unsaid is the possibility of deep differential rotation in the interior of a giant planet which, in the extreme case, does not even admit the notion of a single bulk rotation period. \par

All the above motivates us to describe what we know about the rotation periods and shapes of the icy giants and explore what possibilities exist to possibly reconcile these observations.   

\subsection*{Rotation}
The atmospheres of Uranus and Neptune are typically associated with strong zonal winds with velocities up to 200 m s$^{-1}$ and 400 m s$^{-1}$, respectively. Uranus and Neptune wind velocities are the strongest observed in the solar system; the energy source to drive such energetic atmospheric winds is unknown. 
These wind speeds, however, are relative to assumed solid-body rotation periods of the planets based on Voyager 2 measurements of periodic variations in the planets'  radio signals and fits to the magnetic field observations. The Voyager 2 rotation periods of Uranus and Neptune based on radio data are 17.24h (Desch et al., 1986; Warwick et al.,1986) and 16.11h (Warwick et al., 1989), respectively. The Voyager 2 rotation period of Uranus inferred from fits to the magnetic field data is 17.29h (Ness et al., 1986; Desch et al., 1986) in approximate agreement with the radio period. However, the rms residuals of the Uranian magnetic field fits exhibit a broad minimum in the period ranging from 16.6h to 18h (Desch et al., 1986).
For Neptune no accurate magnetic field period could be derived due to the short Voyager 2 flyby, though a good fit to the magnetic data was obtained for a rotation period of 16h 3m (Ness et al., 1989). \par

\section{Gravitational Fields and Shapes}

The external potential $U$ of a rotating planet can be expressed in a truncated series of even zonal harmonics by (Kaula, 1968), 
\begin{equation}
U = \frac{G M}{r} \left( 1 - \sum_{n=1}^\infty \left( \frac{a}{r} \right)^{2 n} j_{2 n} P_{2 n} \left( \mu  \right)  \right)+ \frac{1}{2} q \left( \frac{G M}{a} \right) \left( \frac{r}{a} \right)^2 (1-\mu^2).
\label{U}
\end{equation}
where $G M$ is the gravitational constant times the total planetary mass, $a$ is the equatorial radius, $j_{2 n}$ are the gravitational coefficients, and $P_{2n}$ are the Legendre polynomials.  The point where the potential is evaluated is given by the planetocentric radius $r$ and latitude $\phi$ ($\mu=sin\phi$).
The second term on the right side is the centrifugal potential. The parameter $q=\omega_0^2a^3$/$GM$, where $\omega_0$ is the angular velocity of rotation. \par

Observations of the motion of planetary rings and satellites and also radio Doppler measurements of the Voyager 2 flybys, have yielded values of $GM$ and the gravitational coefficients $j_2$ and $j_4$ for both Uranus and Neptune. The sixth harmonic $j_6$ has never been detected for either planet. A summary of the values of these quantities can be found on the NASA/JPL website http://ssd.jpl.nasa.gov/?gravity\_fields\_op. Both gravitational fields from this website are given in Table 1. \par

{\bf [Table 1] }\par

For Uranus the occultation of Voyager 2 yielded two radii on ingress and egress. This was a nearly equatorial occultation and it provided essentially a direct measurement of the equatorial radius to $\pm$ 4 km, as given in Table 1. However, there is no measurement of the polar radius of Uranus. Lindal (1992) extrapolated the equatorial radius to the pole using wind velocities (Smith et al. 1989) to obtain the polar radius listed in Table 1 with the larger formal uncertainty of $\pm$ 20 km. \par

The Neptune occultation geometry was not equatorial, and further, only one planetocentric radius measurement was possible, a measurement on egress at a latitude of -42.26$^o$ south. The radius at this latitude is 24601 $\pm$ 4 km. Lindal (1992) extrapolated this single measurement both to the equator and to the south pole, using  Neptune's  wind velocities available at that time (Smith et al. 1989; Hammel et al. 1989), to obtain the equatorial and polar radii given in Table 1; the standard errors of $\pm$ 15 km and $\pm$ 30 km reflect the uncertainties in the extrapolation. \par

The available shape data for both Uranus and Neptune are therefore quite limited and measurements might be even more uncertain due to the challenging orbit determination at the large distances of the outer solar system. We conclude that the shapes of Uranus and Neptune are not well known though the cited shapes listed in Table 1 are prevalent in the literature. Stellar occultations (e.g., French et al., 1998, 1987; Millis et al., 1987), which provide information on the atmospheres of the planets at microbar pressure levels, do not constrain the 1 bar level planetary shapes because of the unknown wind systems between the vastly different pressures.  \par

The solid-body rotation periods based on the Voyager radio periods and listed in Table 1 disagree with the shapes summarized in the same table. This is not surprising since the values of the equatorial and polar radii for Neptune and the polar radius for Uranus have been inferred by Lindal's extrapolation using zonal wind velocities, not the Voyager rotation rate. To demonstrate the point that the commonly used planetary radii of Uranus and Neptune should {\it not} be used with Voyager's radio periods we evaluate equation (1) at both the equator ($\phi=0$, $r=a$) and pole ($\phi=\pi/2$, $r=c$), equate the expressions, solve for the assumed constant $\omega_0$, and convert $\omega_0$ to the period of rotation $\tau_{rot}$ to obtain the rotation period which is associated with the planetary figure by 
\begin{equation}
\tau_{rot} = 2\pi \sqrt{\frac{a^3}{2GM}}\left(\left(\frac{a}{c}\right)-1-j_2\left[\left(\frac{1}{2}\right) + \left(\frac{a}{c}\right)^3 \right] -j_4\left[\left(-\frac{3}{8}\right) + \left(\frac{a}{c}\right)^5 \right]  \right)^{-1/2}.
\end{equation}

In (2), we retain only the gravitational coefficients $j_2$ and $j_4$. The parameter values needed to evaluate the right side of equation (2) for Uranus and Neptune are listed in Table 1. The gravitational coefficients $j_2$ and $j_4$ are modified from the measured $J_2$ and $J_4$ to fit equatorial radii of 25,559 and 24,764 km, for Uranus and Neptune, respectively. The shape-associated rotation period of Uranus is found to be about 15.62h, while the period associated with Neptune's shape is found to be  $\sim$ 16.85h; these shape-associated rotation periods are rather different from the Voyager radio and magnetic periods of both planets. 
It is therefore important to realize that Voyager's rotation periods cannot be used together with the planetary shapes listed in Table 1. It must also be emphasized that the planetary shapes themselves are uncertain. Below, we investigate different approaches to reconcile the shapes and rotation periods of Uranus and Neptune.

\section{Reconciliation of Shapes and Rotation Periods Assuming Solid-Body Rotation}
\subsection{Adapting the Shapes}
In one approach, we assume that Voyager's  rotation periods do represent the rotation periods of the deep interiors of Uranus and Neptune and compute the self-consistent planetary shapes in the absence of distortions by atmospheric winds. We fix the rotation periods of the planets at their measured radio values and determine the geoids that are associated with these rotation rates.  
We derive geoids corresponding to the one-bar equatorial radius of Uranus listed in Table 1 and the observed radius of Neptune at planetocentric latitude -42.26$^o$. These are the only known radii for Uranus and Neptune. The procedure for deriving the geoids is described in detail in Helled et al. (2009a,b). \par

We calculate the geoid radius to fifth order in the small rotation parameter and extrapolate the harmonics $j_2$ and $j_4$ to even harmonics $j_6$, $j_8$ and $j_{10}$ by a linear function in log(j$_{2i}$) versus log(2i). The point here is not to represent higher powers of j$_{2i}$ that have never been measured, but to make sure the precision of the geoid calculation is better than $\pm$ 1.0 km. By symmetry, the functions $r_i(\mu)$ contain only even powers of $\mu$. There is an implicit assumption in the procedure outlined here that the mass in the atmosphere above the one-bar level is negligible. It makes both a negligible contribution to the external potential, and it affects the geoid surface at well under $\pm$ 1.0 km. In the absence of any planetary dynamics, the shape of the planet at 1 bar should conform to this reference geoid. \par

{\bf [Table 2] }\par

The results of the geoid calculations at the 1 bar pressure level in the atmospheres of Uranus and Neptune are given in Table 2. Uranus' equatorial radius is fixed to its measured value (25,559 km). Neptune's equatorial radius is adjusted iteratively until it matches the observed radius of 24,601 km at latitude -42.26$^o$ ($\mu$ of -0.67250). %The sign of $\mu$ is irrelevant in matching Neptune's measured radius. 
Except for the equatorial radius of Uranus, the values of the equatorial and polar radii in Table 2 do not agree with the usual listing of these parameters in compendia of planetary data (Table 1). This is because the extrapolation of the Voyager 2 radio occultation radii to the equator and pole for both Uranus and Neptune was done by Lindal et al. (Lindal et al., 1987; Lindal, 1992) with rotation rates more commensurate with atmospheric wind data from Voyager 2 imaging (Smith et al., 1986, 1989), rather than the radio astronomy periods. 
\par

\subsection{Rotation Periods from Minimization of Dynamical Heights and Wind Speeds}
In this section we determine the solid body rotation periods of Uranus and Neptune that would minimize either the dynamical heights of their isobaric surfaces or their  atmospheric wind speeds. These procedures have been used by Anderson and Schubert (2007) and Helled et al. (2009b) to estimate the rotation periods of Jupiter and Saturn. The methods give a rotation period in good agreement with the radio and magnetic rotation period of Jupiter; the rotation period of Saturn is uncertain. The minimization of winds or dynamical heights is qualitatively plausible but it does not derive from any quantitative argument starting from the equations of motion. That it works for Jupiter is encouraging, but not a proof that it applies to any other planet. It also might work for Saturn since the dynamical heights minimization period is in good agreement with the rotation period inferred by Read et al. (2009) from a dynamically based stability criterion. There is, however, no general acceptance that either method has correctly inferred Saturn's rotation rate. We use the dynamical height and wind minimization methods to investigate possible values of Uranus and Neptune rotation periods without any claim that the values determined from these approaches are the real values for the planets.

\subsection{Dynamical Heights}
The dynamical heights $h(\phi)$ of a pressure isosurface are the altitudes of the surface above the geoid. The dynamical heights therefore depend on the assumed shape of the underlying geoid, which depends on the planetary rotation rate (see details in Helled et al., 2009b; Lindal et al., 1985). Fig. 1 shows the physical shape and dynamical heights of the 1 bar isosurface for both Uranus and Neptune. It should be remembered that these planetary shapes and dynamical heights correspond to the shape parameters listed in Table 1, with the wind velocities relative to Voyager's rotation periods.  
In order to find the rotation periods that minimize the dynamical heights of the 1 bar isosurfaces of Uranus and Neptune without using the extrapolations of Lindal et al. (1987) and Lindal (1992), we use only the measured occultation radii as constraints. The geoids are then constructed by using only the measured radii  ($r=25,559$ km at the equator for Uranus, and $r=24,601$ km at 42.26$^o$S for Neptune), and an assumed solid-body rotation period which we vary. The radii and rotation parameters define a reference geoid. \par 

{\bf [Figure 1] }\par

Dynamical heights $h(\phi)$ with respect to a reference geoid with a constant solid-body rotation period $\omega_0$ are given by (Lindal et al., 1985)
 \begin{equation}
h(\phi) = \frac{1}{g}\int_{\phi}^{\frac{\pi}{2}}V_W\left(2\omega_{0\mathrm{ref}}+\frac{V_W}{r_{\mathrm{ref}}(\phi)cos\phi}\right)\frac{sin(\phi+\psi_{\mathrm{ref}})}{cos\psi_{\mathrm{ref}}}r_{\mathrm{ref}}(\phi)d\phi,
\end{equation}
where $g$ is the average gravity acceleration, $\psi$ is a small angle that gives the difference between the planetocentric latitude $\phi$ and the planetographic latitude $\phi'$, $\psi =\phi'-\phi$, $r$ is the distance to the center of the planet, and $V_W$ is the zonal wind velocity, assumed to be independent of altitude. Parameters with the subscript 'ref' represent quantities calculated for the reference geoid. \par
The wind velocities for Uranus and Neptune are taken from functions that fit the zonal wind data (Sromovsky et al., 1993), 
\begin{eqnarray}
V_{W_U}[m s^{-1}]=27.46 + 36.568cos[\phi] - 175.486cos[3\phi]\\
V_{W_N}[m s^{-1}]=-389 + 0.188 \phi^2 - 
  1.2\times10^{-5}\phi^4,
\end{eqnarray}
where $V_{W_U}$ and $V_{W_N}$ are the wind velocities for Uranus and Neptune, respectively. The above representation for Neptune results in infinite velocities at the poles. We therefore follow French et al. (1998) and apply a cosine taper at high latitudes ($|\phi| > \phi_{max} = 75^o$), so the velocities go to zero at the poles. Since the wind velocities are given with respect to the Voyager rotation periods, the winds are modified to be consistent with the different assumed solid-body rotation rates. We then search for the solid-body rotation periods that minimize the dynamical heights of the 1 bar isosurfaces of Uranus and Neptune. \par

{\bf [Figures 2 and 3] }\par

For Uranus, the solid-body rotation period that minimizes the dynamical heights of its isobaric surface is found to be about 16.58h, with an underlying geoid  polar radius of 25,023 km and equatorial radius of 25,558.6 km. %This result is in agreement with the wind minimization method presented above. 
The altitudes of the 1 bar isosurface above the reference geoid for three different rotation periods  for Uranus, including the Voyager radio period, are presented in Fig. 2. 
The dynamical heights for the Voyager radio period (17.24h) are smaller than the dynamical heights presented in Fig 1. This is due to the fact that the shape of the underlying geoid was defined using Uranus equatorial radius and Voyager's rotation period, while the heights shown in Fig. 1 correspond to an underlying geoid with rotation rate different from Voyager's (Lindal, 1992; Lindal et al., 1987).   
The shape of the 1 bar isosurface for Uranus with a rotation period of $\sim$ 16.58 h is presented in Fig. 3. For comparison, we also present the shape for Voyager's radio rotation period with the geoid shape being determined by the measured equatorial radius (dashed curve). As can be seen from the figure, Uranus' shape with the modified rotation period agrees better with the measured equatorial radius since the dynamical heights are small. \par

{\bf [Figures 4 and 5] }\par

Neptune's dynamical heights are minimized for a solid-body rotation period of $\sim$ 17.46 h. Fig. 4 shows the dynamical heights for three different rotation periods. The dynamical heights for the Voyager rotation period are large in magnitude and negative for Neptune. As a result, a longer rotation period than the radio period is found to minimize the dynamical heights. The corresponding geoid for a rotation period of 17.458h has an equatorial radius of 24,787 km, a polar radius of 24383 km, and a radius of 24,601 km at latitude -42.26$^o$. Fig. 5 shows Neptune's 1 bar isosurface (physical shape) for a rotation period of 17.458h, and the shape for Voyager's rotation period (dashed line), with the constraint on the underlying geoid being $r=24,601$ km at 42.26$^o$S. 
Since the effects of atmospheric dynamics are minimized, the physical shape with the modified period is in good agreement with the measured radius of $r=24,601$ km at 42.26$^o$. \par

For the modified rotation periods, the atmospheric winds on Uranus and Neptune become similar, and have more moderate  speeds. In the next section we show that the rotation periods that minimize the dynamical heights also minimize the wind speeds and we present the revised wind velocities for the modified periods.   

\subsubsection{Wind Speeds} 
It has been demonstrated (Anderson and Schubert, 2007; Helled et al., 2009b) that radio occultation radii and HST wind data for Jupiter and Saturn can be fit to an rms level of ($\sim$ 30 m s$^{-1}$) with a geoid rotating at a uniform angular velocity. For Uranus and Neptune, Voyager's radio rotation periods result in wind profiles with velocities exceeding 200 m s$^{-1}$ for Uranus (Hammel et al., 2001) and 400 m s$^{-1}$ for Neptune (Sromovsky et al., 2001). In this section we search for solid-body rotation periods that minimize the wind velocities on both planets. \par

The wind speeds and rotation periods of features on Uranus at various planetographic latitudes $\phi_g$ are taken from Table II in Hammel et al. (2001). We delete three observations with relatively large standard errors; the Voyager 2 period at -4.5$^o$ (Lindal et al., 1987) and two 1994 HST observations at -24.5$^o$ and -35.0$^o$. This leaves 18 HST observations and 8 observations from Voyager 2 imaging, including one uniquely useful observation at -71.7$^o$ (Smith et al., 1986). Winds on Uranus are found to be minimized for a rotation period of 16h 34m 24s (16.5789h) similar to the rotation period derived from the dynamical height minimization method. This agreement is expected because the dynamical heights are proportional to the wind velocities (see equation (3)), and therefore minimizing the wind speeds is comparable to minimizing the planet's dynamical heights. 
The results for Uranus are shown in Fig. 6 (a). \par

{\bf [Figure 6] }\par

For Neptune, we use the data given by Sromovsky et al. (2001) at various planetocentric latitudes $\phi$. 
The conversion between the two latitude systems is represented to sufficient precision ($<$ 0.1$^o$) by (Stacey, 1992),
\begin{equation}
tan\phi_g=\frac{tan\phi}{(1-f)^2}
\end{equation}
where $f$ is the flattening (a - c)/a. The values of $f$ for Uranus and Neptune are given in Table 1. (Lodders and Fegley, 1998). For Neptune, we fit just the 1998 HST data (Sromovsky et al., 2001) and delete the point at -68.7$^o$ with its relatively large standard error. This leaves a total of 18 observations of wind velocity and rotation period versus planetocentric latitude. The winds on Neptune are found to be minimized for a solid-body rotation period of 17.458h. As for Uranus, the rotation period which minimizes the winds is similar to the one found to minimize the dynamical heights of its isobaric surface. The results for Neptune are shown in Fig. 6 (b). We repeat the wind speed minimization approach using the wind profiles given in equations (4) and (5). The rotation periods that minimize the wind speeds of Uranus and Neptune using these profiles are found to be identical to the ones derived by using the HST data points. Table 3. summarizes the results for the wind minimization calculation for both planets. \par

{\bf [Table 3] }\par

The empirical curves for atmospheric winds presented in Fig. 6 for Uranus and Neptune are rather similar, both suggesting wind velocities of up to 200 m s$^{-1}$. The Neptune winds are not as large as those in Fig. 11 of Sromovsky et al. (2001) because the rotation period for Fig. 6 is different. 
The striking similarity of the latitudinal wind profiles for the two planets could be a result of similar atmospheric dynamics. It is also possible that this similarity is evidence of a deep-seated and similar zonal flow in both planets. We explore this possibility in the following section. 

\section{Differential Rotation on Cylinders and the Oblateness of Uranus and Neptune}

It is possible that Uranus and Neptune do not rotate as solid bodies but instead have different rotation profiles such as differential rotation on cylinders. Other types of differential rotation distributions are possible including differential rotation restricted to a thin spherical shell. However, we restrict our consideration of differential rotation to the model of differential rotation on cylinders and explore the effect of such differential rotation on the planetary shape.  \par
As an alternative to the wind curves of Fig. 6, we assume that both Uranus and Neptune rotate on cylinders about the polar axis. In this model there is no longer a constant or uniform angular velocity $\omega_0$, but instead an angular velocity profile $\omega$ which depends on the distance of a cylindrical shell from the spin axis. The normalized distance $\gamma$ is defined by 
\begin{equation}
\gamma = \frac{r}{a}cos\phi
\end{equation}
and the centrifugal potential $Q$ is not defined by $Q=\frac{1}{2}\omega_0^2r^2cos^2\phi$ as for the solid-body rotation case, but by the integral (Zharkov and Trubitsyn, 1978) 

\begin{equation}
Q = a^2\int_0^\gamma \omega^2\gamma d\gamma.
\end{equation}
In evaluating (8) we consider rotation distributions $\omega(\gamma)$ of the form
\begin{eqnarray}
\omega = \omega(\gamma), 1\ge \gamma \ge \gamma_c\\
\omega = \omega(\gamma_c), 0<\gamma<\gamma_c
\end{eqnarray}
The case of no differential rotation is represented by $\gamma_c=1$, and the case of total differential rotation by $\gamma_c=0$. For $\gamma_c$ near 1 only a thin near equatorial slice of the atmosphere is in differential rotation.  

First, we take $\gamma_c=0$  and investigate models in which the entire planet rotates differentially. The geoid is still defined as before with the total potential $U$ constant on level surfaces. 
To infer the differential rotation profiles $\omega(\gamma)$ we use zonal wind data and find that of all polynomial fits to the wind data for both planets, a simple quadratic in $\gamma$ provides the minimum variance estimate for $\omega^2$. The wind profiles for Uranus and Neptune shown in Fig. 6 can be replaced by a simple differential rotation distribution of the form 
\begin{eqnarray}
\omega^2=\omega_0^2(\alpha_0+\alpha_2\gamma^2)
\end{eqnarray}
with $\omega$ replacing $\omega_0$, the constant angular velocity of rotation (see section 2). As a first guess for $\gamma$  we use the geoids of the previous section and find the best fit to weighted values of $\omega^2$ from the observed periods and their standard errors at various latitudes $\varphi$. We increase the standard error on the Neptune period at latitude 26.8$^o$ from 0.03 hr to 0.48 hr, comparable to other errors at nearby latitudes. Otherwise it dominates the $\chi^2$ statistic for the fit. This procedure yields a geoid based on differential rotation, and the process is iterated by updating $\gamma$ on each iteration. %The geoid under differential rotation converges in one iteration. \par

{\bf [Table 4] }\par

The best-fit parameters for Eq. 11 are given in Table 4. 
The fits to $\omega^2$ are shown in Fig. 7. The two fits are similar, and reflect the similarity of the curves shown in Fig. 6. 
The fits shown in Fig. 7 indicate a large variation in differential rotation throughout the interiors of the two planets. If this is a realistic physical model, it is not a small perturbation to the models assuming uniform rotation. It could have a substantial impact on interior models that fit the observed data for gravitational field, size, and total mass. Finally we compute the geoids by substituting Eq. 11 into Eq. 8 and by using the same procedure outlined earlier (see Helled et al., 2009a,b for details). As with the geoids for solid-body rotation, the gravitational harmonics $J_6$ to $J_{10}$ are extrapolated from the observed J$_2$ and J$_4$. However, for differential rotation, the computed values of these higher harmonics from an interior model are not expected to follow a solid-body extrapolation (Hubbard, 1999). The extrapolation is done simply to have an expression for the radius of the one-bar level surface to fifth order in $q_0$. \par

{\bf [Figure 7] }\par

It has been suggested by Hubbard et al. (1991) that the penetration depth of the differential rotation for Neptune cannot be very deep, a constraint based on gravitational data. Hubbard et al. (1991) have suggested that deep differential rotation results in a positive value for J$_4$, while a negative value is measured. This argument should be relevant to Uranus as well due to its strong resemblance to Neptune. As a result, if differential rotation occurs on Uranus and Neptune, according to Hubbard et al. (1991) it cannot involve a significant amount of mass. We therefore consider other differential rotation distributions that involve only the region of the planets with cylindrical radii that exceed a given value $\gamma_c$. With $\gamma_c$ close to 1 there is little mass in the differentially rotation region. The polar radius of the geoid for these different cases can change by $\sim$ 100 km.  Figures 8 and 9 show the radii and the flattening (f) for different differential rotation distributions of Uranus and Neptune. 
The shape given in Table 1 can be reproduced only with  $\gamma_c=0$; i.e., total differential rotation. As a result we conclude that the radii commonly used for Uranus and Neptune (Table 1; Lindal et al., 1987; Lindal, 1992) can be derived only under the assumption of full differential rotation. \par 

{\bf [Figures 8 and 9] }\par

\section{Summary and Conclusions}
In this paper we use zonal wind data and measured planetary radii to investigate the rotation and shapes of Uranus and Neptune. We point out that the commonly accepted  shapes of Uranus and Neptune should not be used together with Voyager's rotation period, and can be reproduced only with different solid-body rotation periods, or under the assumption of differential rotation of the planets' interiors. We suggest that the solid-body rotation periods based on Voyager radio and magnetic measurements might not represent the rotation periods of Uranus' and Neptune's deep interiors, and investigate different approaches to derive possible rotation periods of these planets.  \par

Internal solid body rotation periods of 16.5789h and 17.4581h are found to minimize the dynamical heights and atmospheric zonal wind velocities of Uranus and Neptune, respectively. These rotation periods are different from the Voyager radio astronomy periods of 17.24h and 16.11h, for Uranus and Neptune, respectively. The derived rotation periods result in very similar winds on Uranus and Neptune with wind velocities of up to 200 meters per seconds for both planets. \par 

The possibility that Uranus and Neptune rotate differentially on cylinders has also been investigated. We show that differential rotation on cylinders derived from wind data can be used to reconstruct Uranus and Neptune measured occultation radii. To reproduce the shapes commonly used for these planets (e.g., Table 1), total differential rotation is required.  In the case of full differential rotation on cylinders there should be no residual atmospheric winds. The models of Uranus and Neptune with fully  differentially rotating interiors and negligible atmospheric winds may, however, be inconsistent with the planets' gravitational data  (Hubbard et al., 1991). \par %As a result, the shapes usually given for Uranus and Neptune, which were derived under the assumption of total differential rotation, are inconsistent with the planets' measured  gravitational fields (jerry does not like this sentence..).  \par

Due to limited data we cannot discriminate among different solid-body rotation periods or the nature of rotation (solid-body vs. differential) for Uranus and Neptune. We point out that the shapes of Uranus and Neptune are not well known. Voyager radio occultations measured only one radius for each planet, giving the equatorial radius of Uranus and the radius of Neptune at 42.26$^o$S. On the basis of our experience with Saturn (Gurnett et al., 2007; Anderson and Schubert, 2007; Helled et al., 2009a), we must consider the possibility that the Voyager radio astronomy and magnetic periods of Uranus and Neptune do not reflect their deep, solid body rotation periods. In addition, radio astronomy and magnetic field periods measure the rotation of the region in which the magnetic field lines are anchored within the planet's interior. Given the unusual magnetic fields of Uranus and Neptune, it is uncertain where the magnetic field lines are actually pinned, and therefore what region is represented by the magnetic measurements. \par

Future investigations of Uranus and Neptune from space will provide more constraints and better understanding of the planets' shapes and rotation rates. Our results may have implications for modeling the interior structures of Uranus and Neptune.  \par

\subsection*{Acknowledgments} 
We thank F. M. Flasar and an anonymous referee for constructive comments. 
R. H. and J. D. A acknowledge support from NASA through the Southwest Research Institute. G. S. acknowledges support from the NASA PGG and PA programs.

\section{References}

Anderson, J. D. and Schubert, G. "Saturns Gravitational Field, Internal Rotation, and Interior Structure." Science 317: (2007) 1384---1387. \\
Desch, M. D., J. E. P. Connerney, and M. L. Kaiser. ÒThe rotation period of Uranus.Ó Nature 322: (1986) 42.\\
French, R. G., Jones, T. J. and Hyland, A. R. "The 1 May 1982 stellar occultation by Uranus and the rings Observations from Mount Stromlo Observatory". Icarus 69: (1987) 499--505.\\
French, R. G., McGhee, C. A. and Sicardy, B., 1998. NeptuneÕs Stratospheric Winds from Three Central Flash Occultations. Icarus, 136, 27--49.\\
Gurnett, D. A., Persoon, A. M., Kurth, W. S., Groene, J. B., Averkamp, T. F., Dougherty, M. K., and
Southwood, D. J. 2007, The Variable Rotation Period of the Inner Region of SaturnÕs Plasma Disk.  Science, 316, 442--445.\\
Gurnett, D. A., Lecacheux, A., Kurth, W. S., Persoon, A. M., Groene, J. B., Lamy, L., Zarka, P. and Carbary, J. F. "Discovery of a north-south asymmetry in Saturn's radio rotation period". GRL (2009), 36, 16102. \\
Hammel, H. B., K. Rages, G. W. Lockwood, E. Karkoschka, and I. de Pater. ÒNew Measurements of the Winds of Uranus.Ó Icarus 153: (2001) 229Ð235. \\
Helled, R., Schubert, G. and Anderson, J. D. ÒJupiter and Saturn Rotation Periods.Ó Planetary and Space Science, 2009(b), 57, 1467--1473. \\
Helled, R., G. Schubert, and J. D. Anderson. ÒEmpirical models of pressure and density in Saturn's interior: Implications for the helium concentration, its depth dependence, and Saturn's precession rateÓ, Icarus, 2009(a), 199, 368--377. \\
Hubbard, W. B., Nellis, W. J., Mitchell, A. C., Holmes, N. C., McCandless, P. C., and
Limaye, S. S. (1991). Interior structure of Neptune - Comparison with Uranus. Science,
253, 648--651.\\
Hubbard, W. B. ÒNOTE: Gravitational Signature of JupiterÕs Deep Zonal Flows.Ó Icarus 137: (1999) 357Ð359. \\
Ingersoll, A. P., 1990. Atmospheric dynamics of the outer planets. Science 248, 308Ð315.\\
Jacobson, R. A. 2003, JUP230 orbit solution, http://ssd.jpl.nasa.gov/?gravity\_fields\_op \\
Jacobson, R. A., Antreasian, P. G., Bordi, J. J., Criddle, K. E., Ionasescu, R., Jones, J. B., Mackenzie, R. A.,
Meek, M. C., Parcher, D., Pelletier, F. J., Owen, Jr., W. M., Roth, D. C., Roundhill, I. M., and Stauch,
J. R. 2006, The Gravity Field of the Saturnian System from Satellite Observations and Spacecraft Tracking Data. ApJ, 132, 2520--2526\\
Millis, R. L., Wasserman, L. H. and French, R. G. "Observations of the 22 April 1982 stellar occultation by Uranus and the rings". Icarus 69 (1987) 176--184.\\
Kaula, William M. An Introduction to Planetary Physics: The Terrestrial Planets. New York: John Wiley and Sons, 1968.\\
Kurth,W. S., T. F. Averkamp, D. A. Gurnett, J. B. Groene, and A. Lecacheux, 2008. An update to a Saturnian longitude system based on kilometric radio emissions. J. Geophys. Res., 113, A05222.1--A05222.10\\
Lindal, G. F. ÒThe atmosphere of Neptune - an analysis of radio occultation data acquired with Voyager 2.Ó Astron. Jour. 103: (1992) 967Ð982. \\
Lindal, G. F., Sweetnam, D. N., and Eshleman, V. R. 1985, The atmosphere of Saturn - an analysis of the Voyager radio occultation measurements. AJ, 90, 1136--1146.\\
Lindal, G. F., J. R. Lyons, D. N. Sweetnam, V. R. Eshleman, and D. P. Hinson. ÒThe atmosphere of Uranus - Results of radio occultation measurements with Voyager 2.Ó Jour. Geophys. Res. 92: (1987) 14,987Ð15,001.\\
Lodders, Katharina, and Bruce Fegley, Jr. The Planetary Scientists Companion. New York Oxford: Oxford University Press, 1998. \\
Smith, B. A., L. A. Soderblom, D. Banfield, C. Barnet, R. F. Beebe, A. T. Bazilevskii, K. Bollinger, J. M. Boyce, G. A. Briggs, and A. Brahic. ÒVoyager 2 at Neptune - Imaging science results.Ó Science 246: (1989) 1422Ð1449. \\
Smith, B. A., L. A. Soderblom, R. Beebe, D. Bliss, R. H. Brown, S. A. Collins, J. M. Boyce, G. A. Briggs, A. Brahic, J. N. Cuzzi, and D. Morrison. ÒVoyager 2 in the Uranian system - Imaging science results.Ó Science 233: (1986) 43Ð64. \\
%Soffel, Michael H. Relativity in Astrometry, Celestial Mechanics and Geodesy. Berlin Heidelberg New York: Springer-Verlag, 1989. \\
Sromovsky, L. A., P. M. Fry, T. E. Dowling, K. H. Baines, and S. S. Limaye. ÒNeptuneÕs Atmospheric Circulation and Cloud Morphology: Changes Revealed by 1998 HST Imaging.Ó Icarus 150: (2001) 244Ð260.\\
Stacey, Frank D. Physics of the Earth. Brisbane: Brookfield Press, 1992. \\
Warwick, J. W., D. R. Evans, G. R. Peltzer, R. G. Peltzer, J. H. Romig, C. B. Sawyer, A. C. Riddle, A. E. Schweitzer, M. D. Desch, and M. L. Kaiser. ÒVoyager planetary radio astronomy at Neptune.Ó Science 246: (1989) 1498Ð1501. \\
Zharkov, V. N., and V.P. Trubitsyn. Physics of Planetary Interiors. Tucson: Pachart, 1978. 

\clearpage

\begin{table}[h!]
\begin{center}
\begin{tabular}{lcccl}
\multicolumn{3}{c}{} \\
\hline
\hline
Parameter & Uranus & Neptune\\
\hline
P (rotation period) & 17.24h & 16.11h\\
GM (km$^3$ s$^{-2}$) & 5,793,964 $\pm$ 6& 6,835,100. $\pm$ 10 \\
R$_{ref}$ (km) & 26,200 & 25,225 \\
J$_2$ ($\times$10$^{6}$) &3341.29 $\pm$ 0.72&3408.43 $\pm$ 4.50\\
J$_4$ ($\times$10$^{6}$) & -30.44 $\pm$ 1.02& -33.40 $\pm$ 2.90\\
a (km)  & 25,559 $\pm$ 4 & 24,764 $\pm$ 15  \\
c (km) & 24,973 $\pm$ 20 & 24,341 $\pm$ 30  \\
f (flattening) & 0.02293 $\pm$ 0.0008 &  0.0171 $\pm$ 0.0014\\
\hline
\end{tabular}
	\caption{{\small Physical data, taken from JPL database:  http://ssd.jpl.nasa.gov, Jacobson (2003), Jacobson et al. (2006), and Lindal et al., (1992; 1985). R$_{ref}$ is the reference equatorial radius in respect to the measured gravitational harmonics J$_2$ and J$_4$, $a$ and $c$ are the equatorial and polar radii, respectively.
}}
\end{center}
\end{table}

\begin{table}[h!]
\begin{center}
\begin{tabular}{lcccl}
\multicolumn{3}{c}{} \\
\hline
\hline
Parameter & Uranus & Neptune\\
\hline
P (h) & 17.24 $\pm$ 0.01& 16.11 $\pm$ 0.05 \\
a (km)  & 25,559 $\pm$ 4 & 24,810 $\pm$ 20  \\
c (km) & 25,052 $\pm$ 20 & 24,357 $\pm$ 30  \\
q & 0.029535 & 0.026223  \\
j$_2$ & 0.118876 & 0.134364  \\
j$_4$ & -0.038530 &  -0.051904 \\
j$_6$ & 0.019934 &  0.029755 \\
j$_8$& -0.012489 & -0.020050 \\
j$_{10}$& 0.008690 &  0.014762\\
\hline
\end{tabular}
	\caption{{\small  Results for solid-body rotation at the radio astronomy periods P for Uranus and Neptune. The equatorial radius a and polar radius c are given for the one-bar pressure level in the atmosphere. 
}}
\end{center}
\end{table}

\begin{table}[h!]
\begin{center}
\begin{tabular}{lcccl}
\multicolumn{3}{c}{} \\
\hline
\hline
Parameter & Uranus & Neptune\\
\hline
P (h) & 16.5789 $\pm$ 0.0002 & 17.4581 $\pm$ 0.0007 \\
a (km)  & 25,559 $\pm$ 4 & 24,787 $\pm$ 4  \\
c (km) & 25,023 $\pm$ 4 & 24,383 $\pm$ 4  \\
f (flattening) & 0.2097 & 0.1629 \\
\hline
\end{tabular}
	\caption{{\small  Results for solid-body rotation.}}
\end{center}
\end{table}

\begin{table}[h!]
\begin{center}
\begin{tabular}{lcccl}
\multicolumn{3}{c}{} \\
\hline
\hline
Parameter & Uranus & Neptune\\
\hline
%P (h) & 16.5789 $\pm$ 0.0002 & 17.4581 $\pm$ 0.0007 \\
a (km)  & 25,559 $\pm$ 4 & 24,780 $\pm$ 4  \\
c (km) & 24,976 $\pm$ 4 & 24,351 $\pm$ 4  \\
R & 25,377 $\pm$ 4 & 24,646 $\pm$ 4  \\
q$_0$ & 0.029535 & 0.026223  \\
$\omega_0^2$ ($10^{-8}$ s$^{-2})$& 1.108177 & 0.999449  \\
$\alpha_0$ & 1.3975 $\pm$ 0.0016 & 1.3689 $\pm$ 0.0062  \\
$\alpha_2$ & -0.5584 $\pm$ 0.0022 & -0.5497$\pm$ 0.0090  \\

\hline
\end{tabular}
	\caption{{\small  Results for differential rotation. Adopted parameters for the centrifugal potential Q are listed with no uncertainty. The mean radius R is the radius of an equivalent sphere with the mean density of the planet.  
}}
\end{center}
\end{table}

\clearpage
\begin{figure}[h!]
   \centering
    \includegraphics[width=6.4in]{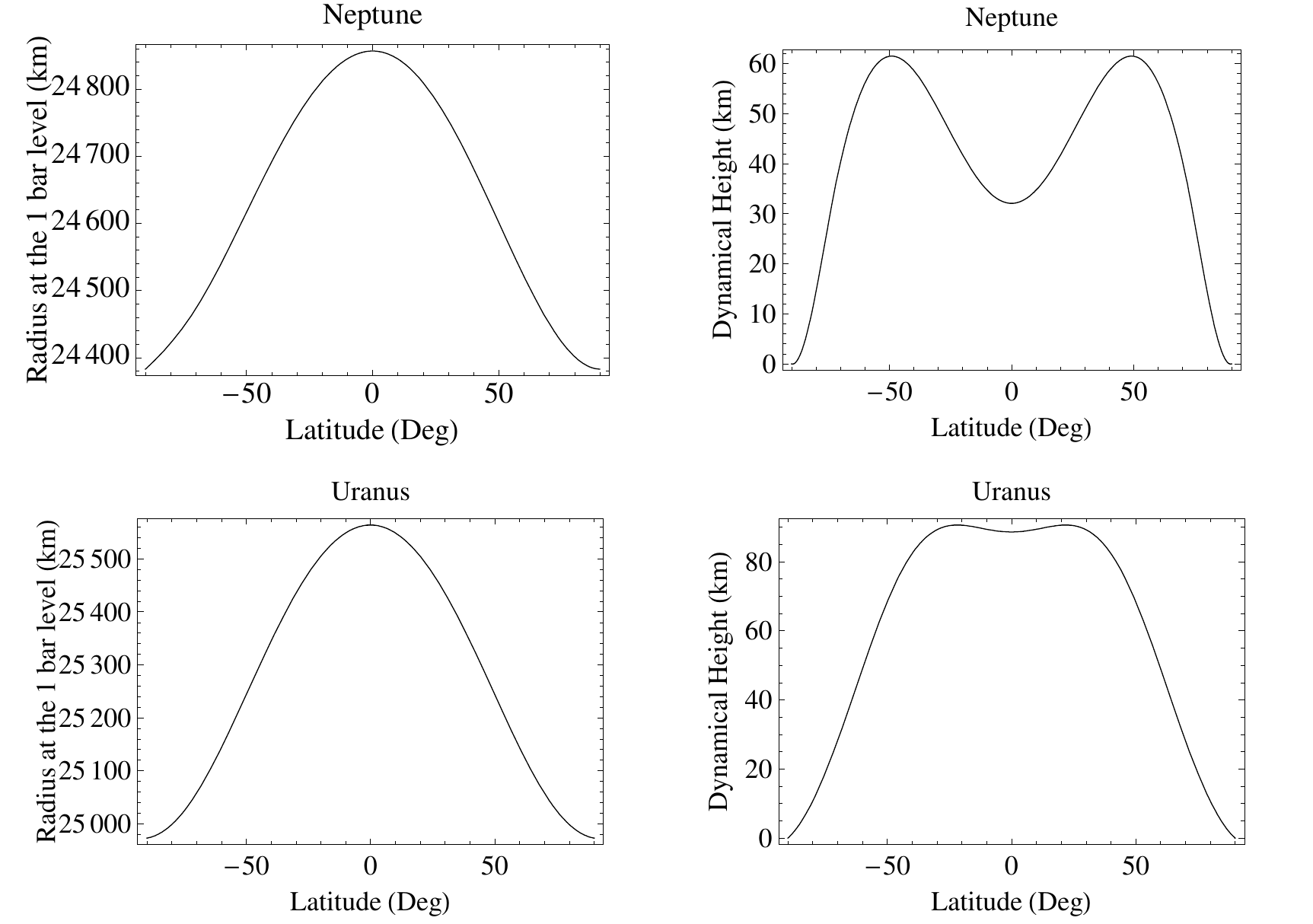}
    \caption[err]{Shapes and dynamical heights of Uranus and Neptune for Voyager rotation periods and planetary shapes given in Table 1. The derived dynamical heights are in agreement with previous work (French et al., 1998, Lindal, 1987).}
\end{figure}

\clearpage
\begin{figure}[h!]
   \centering
    \includegraphics[width=4.0in]{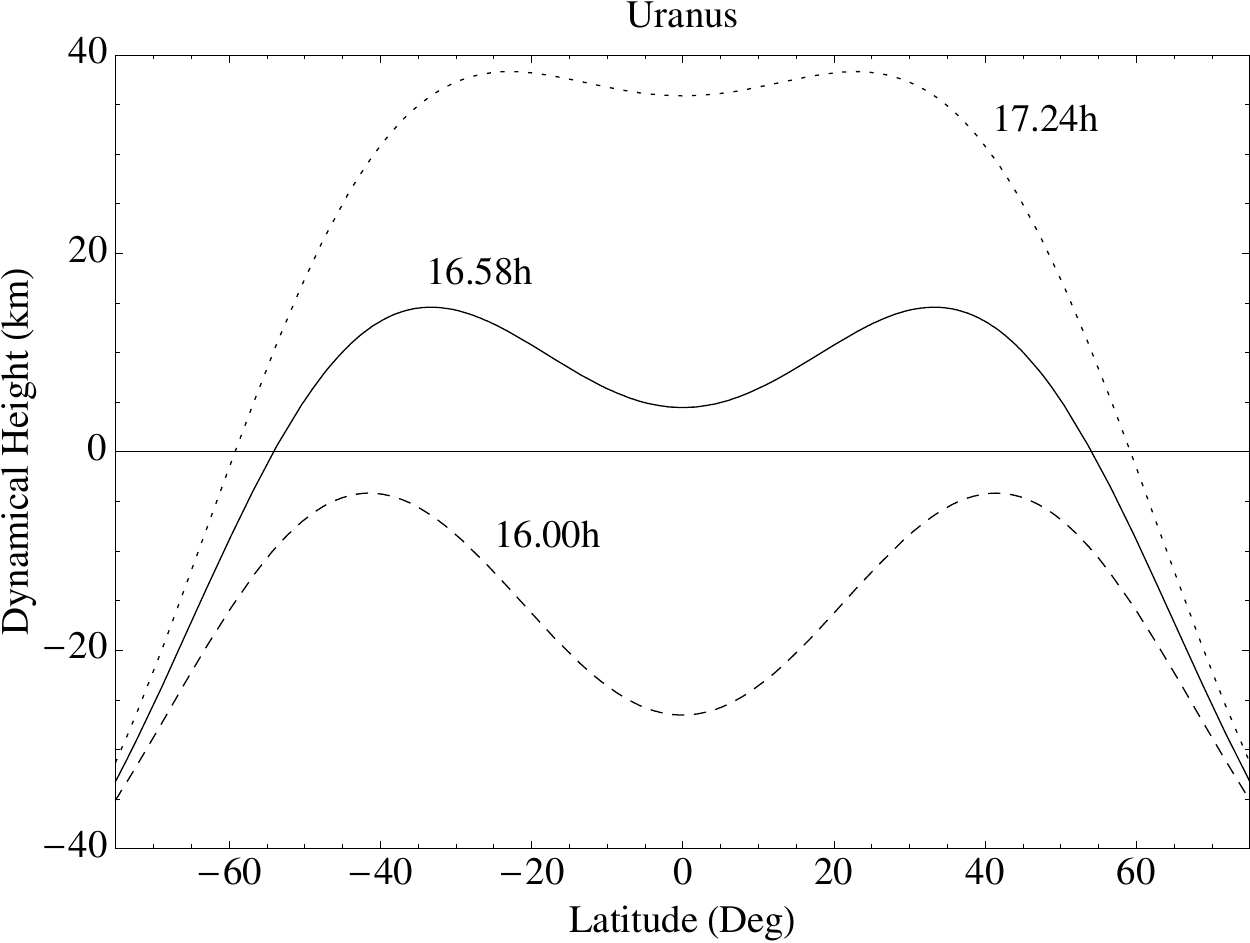}
    \caption[err]{Altitudes of the 1 bar Uranus isosurface above a reference geoid for three different rotation periods. The horizontal line represents a reference geoid unperturbed by zonal winds.}
\end{figure}

%\clearpage
\begin{figure}[h!]
   \centering
    \includegraphics[width=4.0in]{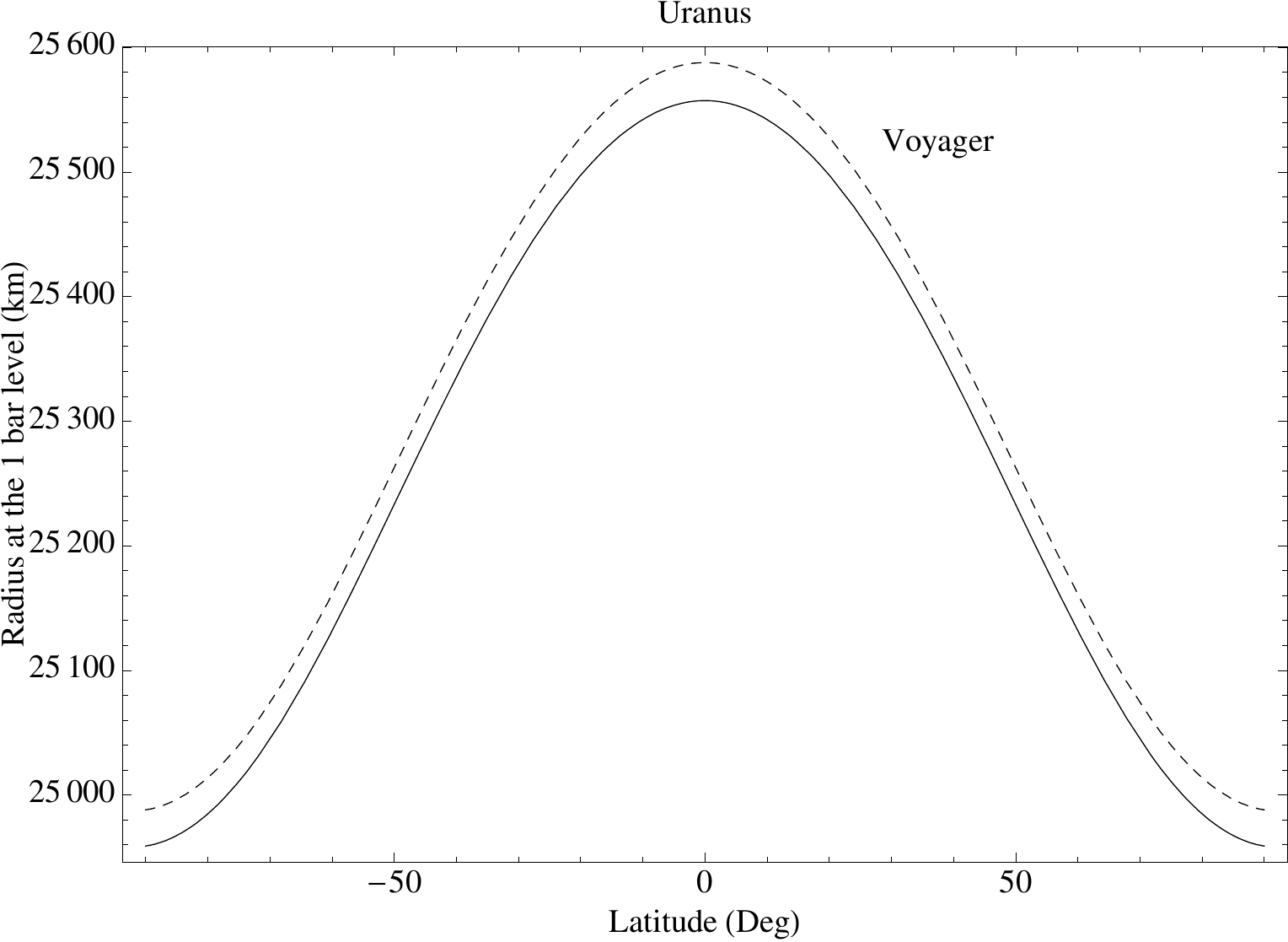}
    \caption[err]{Uranus' physical shape for rotation period of 16.58h, the rotation period that minimizes it's winds and dynamical heights. The dashed line corresponds to the Voyager rotation period of 17.24 h.}
\end{figure}

\clearpage
\begin{figure}[h!]
   \centering
    \includegraphics[width=4.0in]{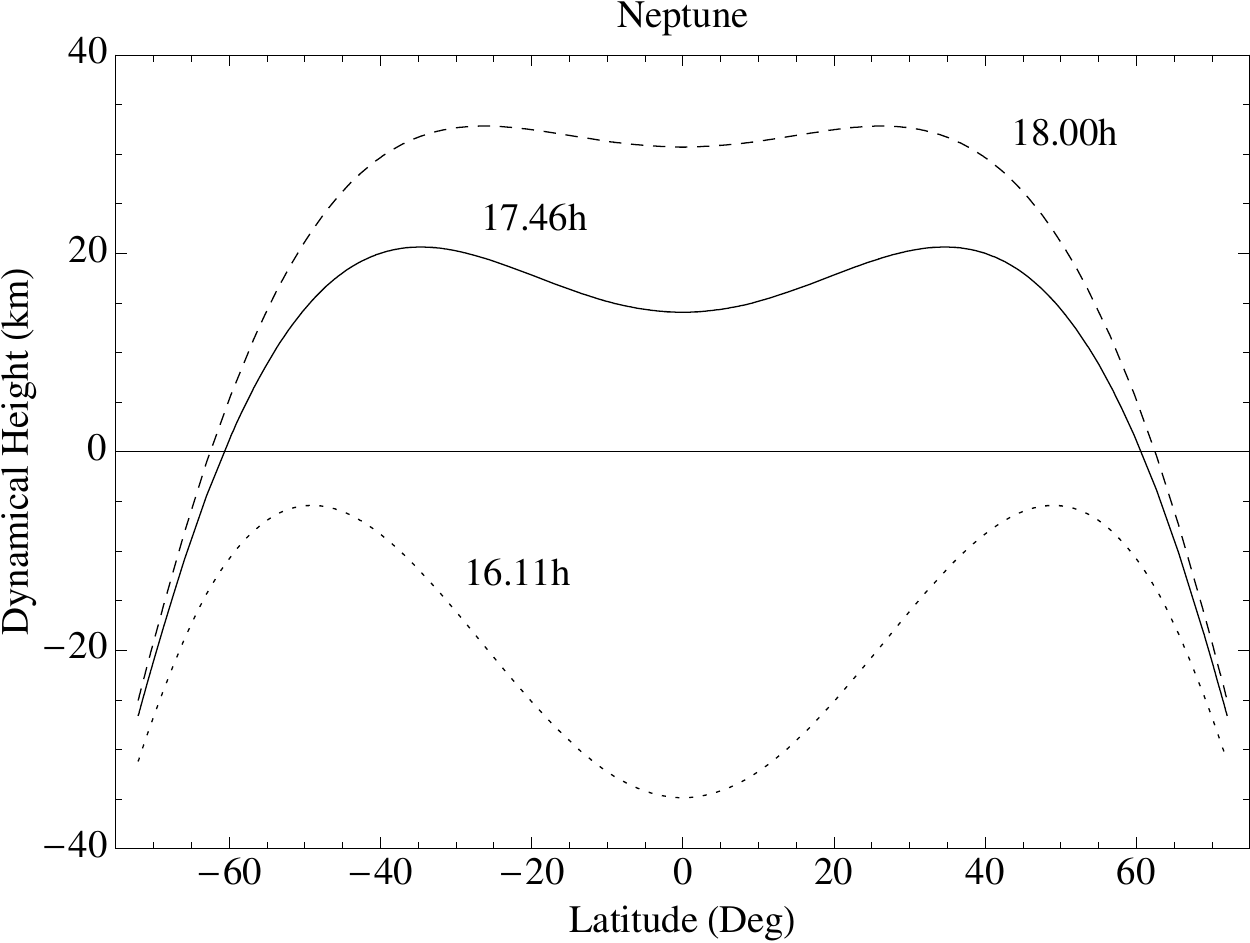}
    \caption[err]{Neptune's altitudes of the 1 bar isosurface above a reference geoid for three different rotation periods. The horizontal line represents a reference geoid unperturbed by zonal winds.}
\end{figure}

\begin{figure}[h!]
   \centering
    \includegraphics[width=4.0in]{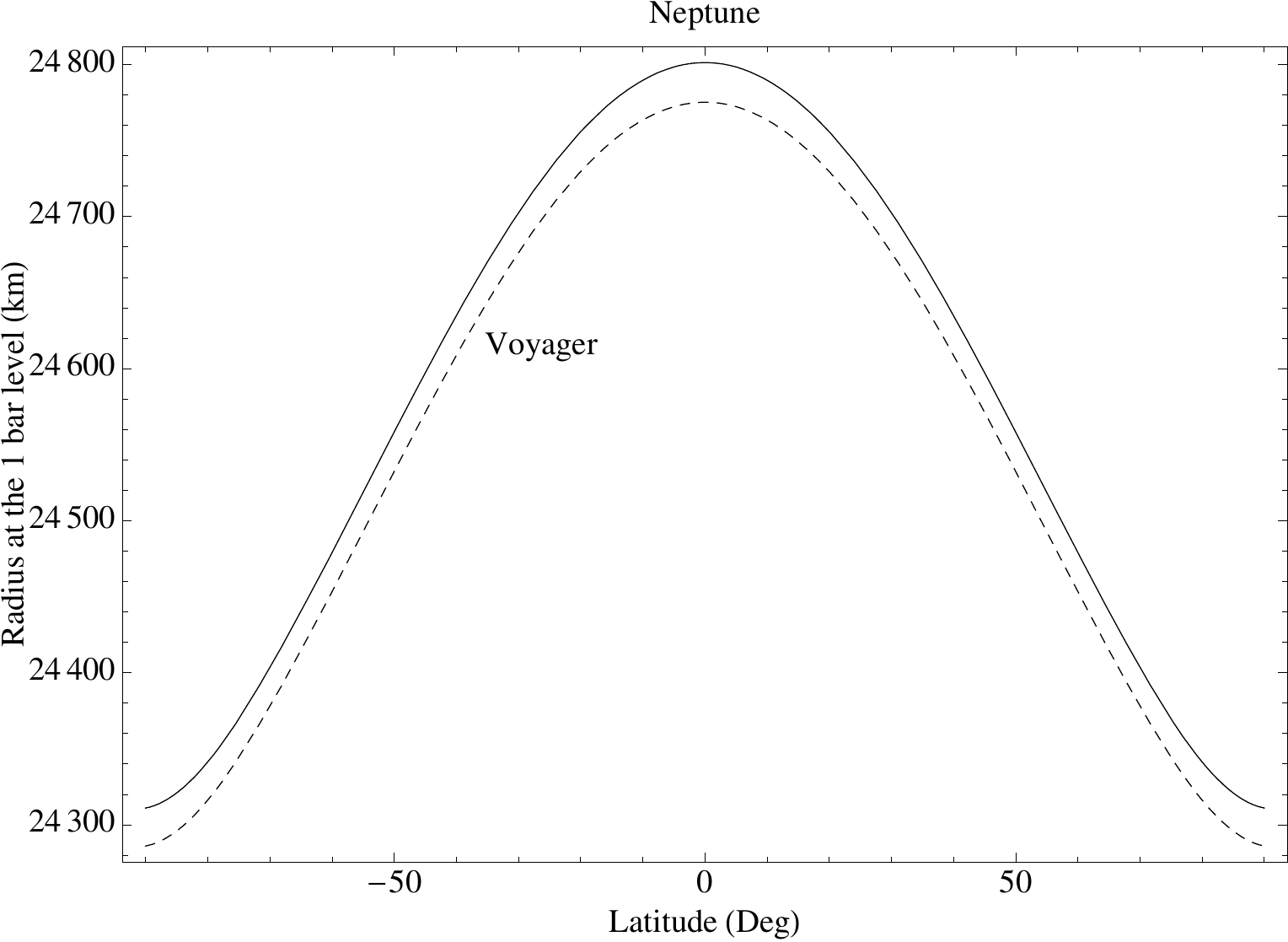}
    \caption[err]{Neptune's physical shape for a rotation period of 17.46, the rotation period that minimizes it's winds and dynamical heights. The dashed line corresponds to the Voyager rotation period of 16.11 h.}
\end{figure}

\begin{figure}
% \vspace*{-2.0 cm}
\begin{center}
 \includegraphics[width=6.20in]{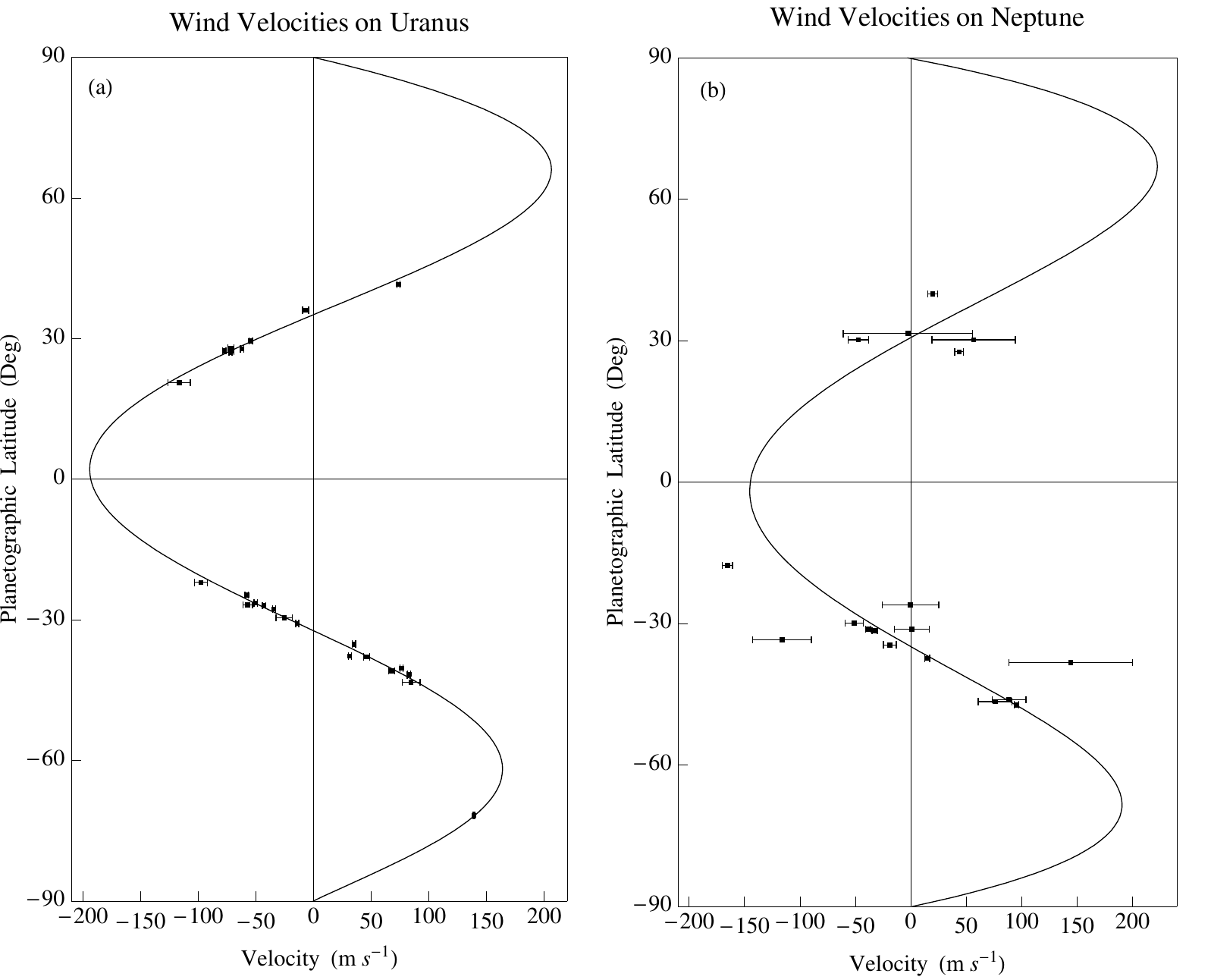} 
% \vspace*{-1.0 cm}
\caption{Best-fit geoids to Uranus and Neptune wind data. (a) Uranus: the rotation period is (16.5789 $\pm$ 0.0002) hr and the adopted equatorial radius is 25,559 km. The corresponding polar radius is (25023 $\pm$ 4) km. The solid line represents a polynomial fit to the wind-velocity residuals referenced to the uniformly rotating geoid. A lack of an error bar for some of the points means that the error in the wind velocity is smaller than the resolution of the plot. (b) Neptune: the rotation period is (17.4581 $\pm$ 0.0007) h. The corresponding equatorial and polar radii are  24,825 $\pm$ 4 km and 24,419 $\pm$ 4 km, respectively, and the corresponding radius at -42.26$^\circ$ is 24,638 km. The solid line represents an empirical polynomial fit to the wind-velocity residuals referenced to the uniformly rotating geoid.}
   \label{UranusSB}
\end{center}
\end{figure}
%**************************************************

%*****************************************
\begin{figure}
% \vspace*{-2.0 cm}
\begin{center}
 \includegraphics[width=5.0in]{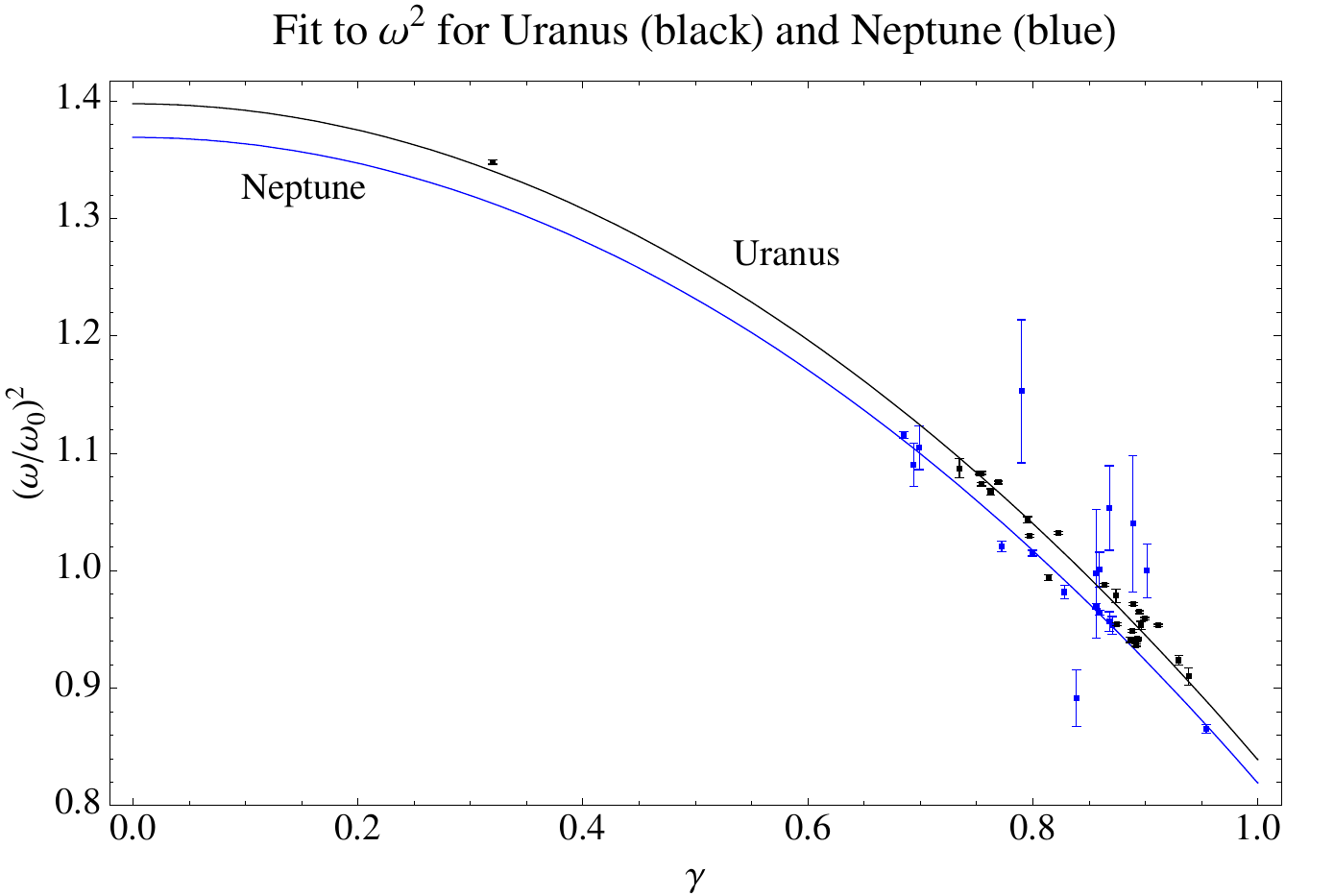} 
% \vspace*{-1.0 cm}
\caption{The parameters for the fits are given in Table~3. For Uranus (black curve) the data for $\omega^2$ are from the observed periods and standard errors as a function of planetocentric latitude $\phi$ (Hammel et al., 2001). For Neptune (blue curve), the data for $\omega^2$ are from Sromovsky et al.~(2001).}
   \label{om2U}
\end{center}
\end{figure}
%**************************************************

%*****************************************

\begin{figure}
% \vspace*{-2.0 cm}
\begin{center}
 \includegraphics[width=4.3in]{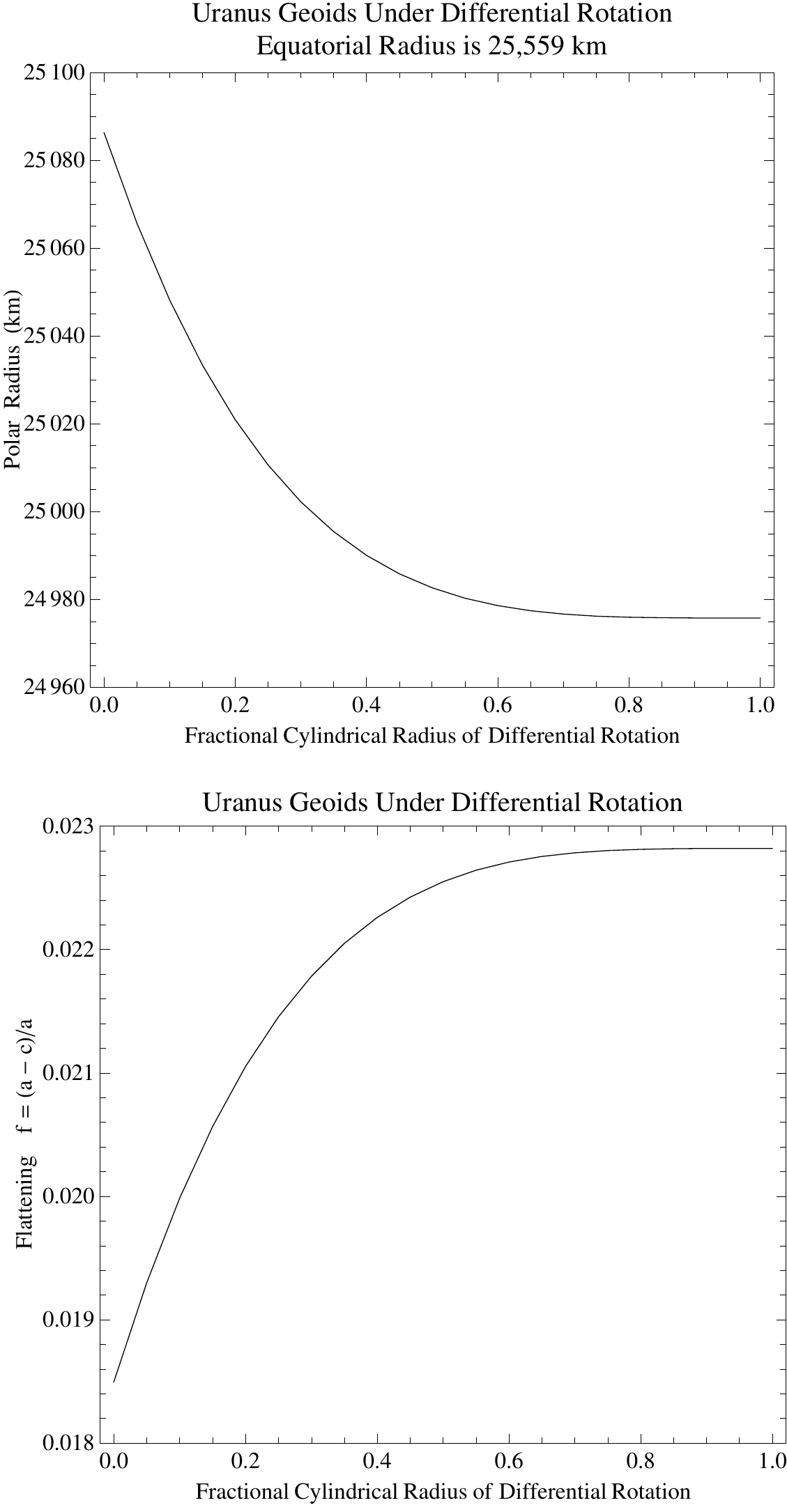} 
% \vspace*{-1.0 cm}
\caption{Uranus Geoids Under Differential Rotation.}
  \end{center}
\end{figure}

\begin{figure}
% \vspace*{-2.0 cm}
\begin{center}
 \includegraphics[width=2.9in]{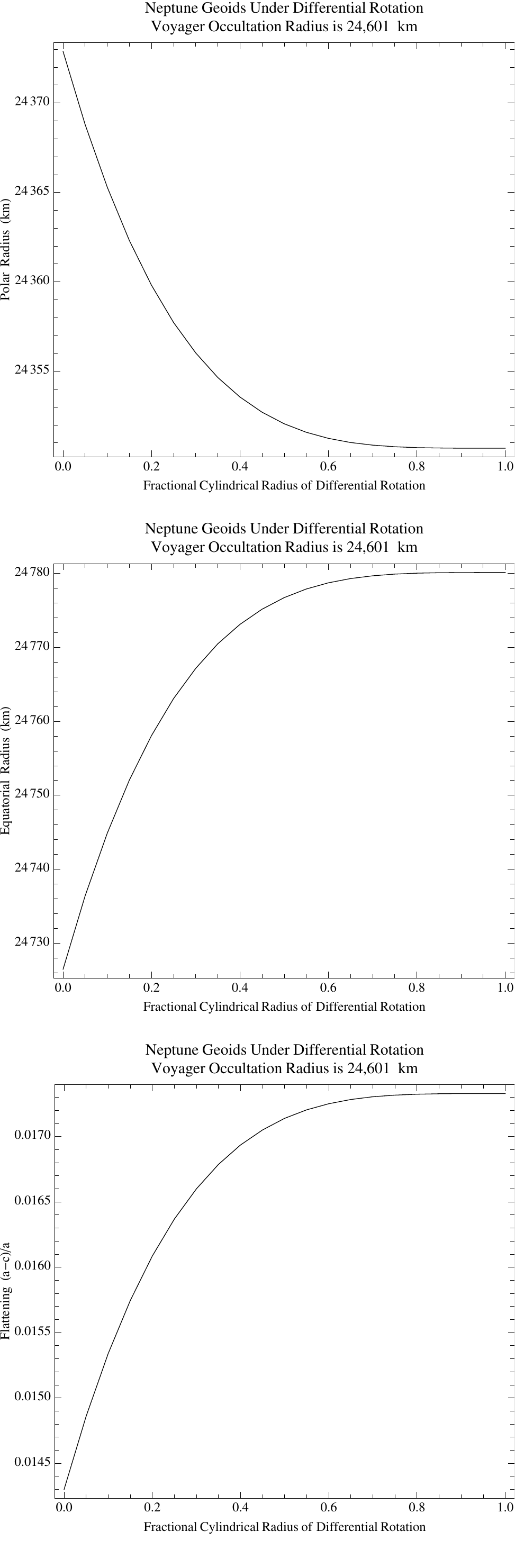} 
% \vspace*{-1.0 cm}
\caption{Neptune Geoids Under Differential Rotation
Voyager Occultation Radius is 24,601 km.}
  \end{center}
\end{figure}

\end{document}